\documentclass[5p,fleqn,sort&compress]{elsarticle}
\usepackage[english]{babel}
\usepackage{hyphenat}
\usepackage[utf8]{inputenc}	%Hvis du benytter windows i stedet for linux, så skift utf8 ud med latin1. Tillader danske tegn.
\usepackage{savesym}
\usepackage{amsmath}
\savesymbol{iint}
\usepackage{txfonts}
\restoresymbol{TXF}{iint}
\usepackage[T1]{fontenc}	%Tillader danske tegn
\usepackage{graphicx}
\usepackage{siunitx}	%Bruges \SI{<tal>}{<enhed>}, \si{<enhed>} eller \num{<tal>}.
\usepackage{url}		 %bruges til at formattere url'er... kan sagtens udelades.
\addto\captionsenglish{\renewcommand{\figurename}{Fig.}}
\newcommand{\tref}[1]{\tablename~\ref{#1}}
\newcommand{\fref}[1]{\figurename~\ref{#1}}

\usepackage{hyperref}
\hypersetup
{   pdfsubject={Nuclear Phenomenology},
	pdfauthor={Jonas Refsgaard}
    pdftitle={Three-body effects in the Hoyle-state decay},
    pdfstartview=FitH,
    colorlinks=true}
    
\makeatletter
\begingroup
\catcode`\_=\active
\protected\gdef_{\@ifnextchar|\subtextup\sb}
\endgroup
\def\subtextup|#1|{\sb{\textup{#1}}}
\AtBeginDocument{\catcode`\_=12 \mathcode`\_=32768 }
\makeatother

\usepackage[]{csquotes} %Danske citationstegn. \enquote{}
\usepackage{bm}
\usepackage{amssymb}

\newcommand{\carbon}{$^{12}\mathrm{C}$}

\newcommand\T{\rule{0pt}{2.3ex}}       % Top strut
\newcommand\B{\rule[-1.0ex]{0pt}{0pt}} % Bottom strut

\begin{document}

%\title{The \texorpdfstring{$3\alpha$}{triple-alpha} decay of the Hoyle state in \texorpdfstring{$^{12}\mathrm{C}$}{12C}; sequential or direct?}
%\title{Final-state Coulomb interactions in the decay of the Hoyle state}
\title{Three-body effects in the Hoyle-state decay}

\author[au]{J.~Refsgaard \corref{cor} \fnref{fn1}}
\ead{jr@phys.au.dk}

\author[au]{H.\,O.\,U.~Fynbo}
\author[au]{O.\,S.~Kirsebom}
\author[au]{K.~Riisager}
\cortext[cor]{Corresponding author}
\address[au]{Department of Physics and Astronomy, Aarhus University, DK-8000 Aarhus, Denmark}
\fntext[fn1]{Present address: Instituut voor Kern- en Stralingsfysica, KU Leuven, B-3001 Leuven, Belgium}

\begin{keyword}
Collective levels \sep breakup and momentum distributions \sep multifragment emission and correlations \sep stellar helium burning

\PACS 21.10.Re \sep 25.60.Gc \sep 25.70.Pq \sep 26.20.Fj
\end{keyword}

\begin{abstract}
We use a sequential $R$-matrix model to describe the breakup of the Hoyle state into three $\alpha$ particles via the ground state of $^8\mathrm{Be}$. It is shown that even in a sequential picture, features resembling a direct breakup branch appear in the phase-space distribution of the $\alpha$ particles. We construct a toy model to describe the Coulomb interaction in the three-body final state and its effects on the decay spectrum are investigated. The framework is also used to predict the phase-space distribution of the $\alpha$ particles emitted in a direct breakup of the Hoyle state and the possibility of interference between a direct and sequential branch is discussed. Our numerical results are compared to the current upper limit on the direct decay branch determined in recent experiments.
\end{abstract}
\maketitle

\section{Introduction}
\noindent In the beginning af the 1950s the stellar triple-$\alpha$ process was proposed as the production mechanism for $^{12}\mathrm{C}$~\cite{Opik1951,Salpeter1952}. A main role in this process is played by the first excited $0^+$ state in $^{12}\mathrm{C}$, also known as the Hoyle state~\cite{Hoyle1953}. It was early realised that the Hoyle state might have a peculiar structure, strongly influenced by $\alpha$ particle clusterisation~\cite{Morinaga1956}.

Studying the decay of the Hoyle state to the $3\alpha$ continuum is one of the methods that has been used to probe its structure. In particular it has been shown that the ratio between its probability for decaying directly to the $3\alpha$ continuum vs. the probability for sequential decay through the $^8\mathrm{Be}$ ground state has an impact on the calculated production rate for $^{12}\mathrm{C}$ in stellar environments~\cite{Ogata2009,Garrido2011,Ishikawa2013}. Consequently, the three-body breakup of the Hoyle state and the phase-space distribution of the emitted $\alpha$ particles have been the subject of an extended experimental campaign, stretching over the past twenty-five years~\cite{Freer1994,Raduta2011,Manfredi2012,Kirsebom2012,Rana2013,Itoh2014,Smith2017,DellAquila2017}. As a result, upper bounds on the direct decay branch have been obtained, the most recent, and also most restrictive, limits being \num{4.7e-4}~\cite{Smith2017} and \num{4.2e-4}~\cite{DellAquila2017} at \SI{95}{\percent} confidence level. In contrast to these results are a couple of measurements that give non-zero values of the direct decay branch, namely \cite{Raduta2011} and \cite{Rana2013}, which put the direct decay branch at \num{1.7 \pm 0.5e-1} and \num{9.1 \pm 1.4e-3}, respectively.

The results are based on particular models for the sequential and direct decays: The sequential branch is modelled using a $\delta$ function to describe the $^8\mathrm{Be}$ resonance. This approach ignores the freedom to populate the $^8\mathrm{Be}$ system also off-resonance and a significant portion of the three-body phase space is thereby excluded. The direct decay is assumed to be a uniform phase-space decay, an assumption which is not taking the Coulomb interaction between the $\alpha$ particles into account. Because the Hoyle state decays through emission of low-energy $\alpha$ particles, far below the Coulomb barrier, Coulomb effects should heavily influence the phase space distribution in a direct decay.

%In several of the analyses some particular forms of the direct decay are used, the three most popular ones being the uniform phase space decay, a decay with equal sharing of energy between the fragments and a decay where two $\alpha$ particles are emitted in opposite directions, leaving the third at rest in the center of mass of the decaying system. None of these models take Coulomb interactions into account in any way. Because the Hoyle state decays through emission of low-energy $\alpha$ particles, far below the Coulomb barrier, Coulomb effects should heavily influence the phase space distribution in a direct decay.

% Since for the Hoyle state decay we are considering the emission of low-energy $\alpha$ particles, far below the Coulomb barrier, we should expect the Coulomb interaction to heavily influence the phase space distribution in a direct decay.

In this paper we employ a sequential $R$-matrix model to address the shortcomings of the simpler models. The main justification for using the sequential model is that it, in several cases, has been shown to describe three-body decays at least as well as more sophisticated theoretical calculations~\cite{Fynbo2003,Kirsebom2010}. We develop a toy model of the final-state Coulomb interaction and show that three-body effects are important for our interpretation of the decay spectrum of the Hoyle state. Furthermore we use the model to mock up the decay spectrum of a hypothetical direct decay and discuss the possibility of interference between sequential and direct decay channels.

%In this paper we present a concrete, phenomenological model that includes the most important physical ingredients of the three-body Coulomb problem. We investigate the predictions of the model and compare to the available experimental data. One of our motivations is to contribute to a discussion on how to treat the Hoyle state decay more systematically than has previously been done.

%In this paper we apply a sequential model that relies on the well-established $R$-matrix description of the intermediate $^8\mathrm{Be}$ system to the decay of the Hoyle state~\cite{Wigner1947,Lane1958}. Our aim is to investigate the predictions of the model and compare to the available experimental data. Furthermore we hope to initiate a discussion on how to treat the direct decay more systematically than has previously been done and whether we can predict the effect of Coulomb interactions in the final state.

\section{The sequential model}
\noindent It is possible to regard the three-body decay of $^{12}\mathrm{C}^*$ as either direct or sequential, by which we mean
\begin{align*}
&^{12}\mathrm{C}^* \rightarrow \alpha + \alpha + \alpha \quad &&\text{(Direct)} \\
&^{12}\mathrm{C}^* \rightarrow {^8\mathrm{Be}}^* + \alpha \rightarrow \alpha + \alpha + \alpha \quad &&\text{(Sequential)}. 
\end{align*}
The sequential interpretation was proposed in 1936 in order to explain the angular and energy distributions of $\alpha$ particles from the ${^{11}\mathrm{B}}(p,3\alpha)$ reaction observed in early cloud chamber experiments. In the sequential picture, the dynamics of the breakup are determined by the properties of the intermediate nucleus, and the $\alpha$-particle distributions were used to deduce the energies and widths of the lowest states of the unstable $^8\mathrm{Be}$ nucleus~\cite{Dee1936,Bethe1937,Wheeler1941}. Later, several theoretical frameworks appeared that could be used to analyse the three-body breakup as a sequential process~\cite{Watson1952,Migdal1955,Phillips1960,Duck1964,Schaefer1970}.

\subsection{General formalism}
\noindent We use a sequential model to calculate the expected phase-space distribution of the $\alpha$ particles emitted from the unstable Hoyle state in {\carbon}. For a particular permutation of the $\alpha$ particles, the decay amplitude is given by
\begin{align}
\label{eq:single_amp}
f_c^{m_a}&(123) = \sum_{m_b} \langle J_b l_1 m_b (m_a-m_b) \vert J_a m_a \rangle \nonumber \\ &\times \bigl[i^{l_1} Y_{l_1}^{m_a-m_b}(\Omega_1)\bigr] \bigl[i^{l_2} Y_{l_2}^{m_b}(\Omega_{23})\bigr]  \nonumber \\ &\times  \gamma_c \bigl(2P_{l_1} / \rho_1\bigr)^{\frac{1}{2}}\exp\bigl[i(\omega_{l_1} - \phi_{l_1})\bigr] F_c(E_{23}),
\end{align}
where $F_c(E_{23})$ is a factor describing the resonant strength of the intermediate system. In the single-level approximation we have
\begin{align}
\label{eq:single_level}
F_c(E_{23}) = \frac{\gamma_{\lambda_b l_2}\bigl(2P_{l_2} / \rho_{23}\bigr)^{\frac{1}{2}}\exp\bigl[i(\omega_{l_2} - \phi_{l_2})\bigr]}{E_{\lambda_b} - E_{23} - \bigl[S_{l_2} - B_{l_2} + iP_{l_2}\bigr] \gamma_{\lambda_b l_2}^2} .
\end{align}
To obtain the total decay weight the expression is symmetrised in the permutation of the $\alpha$ particles:
\begin{align}
\label{eq:total_weight}
W = \sum_{m_a} \Bigl\lvert \sum_c \Bigl\lbrace f_c^{m_a}(123) + f_c^{m_a}(231) + f_c^{m_a}(312) \Bigr\rbrace \Bigr\rvert^2 .
\end{align}
The various symbols appearing in eqs. \eqref{eq:single_amp}--\eqref{eq:total_weight} are explained in \tref{tab:notation}. When we later use eq. \eqref{eq:total_weight} to calculate decay weights, we refer to it as \emph{Model I}.
\begin{table}[htbp]
\centering
\caption{Explanation of the parameters appearing in eqs. \eqref{eq:single_amp}--\eqref{eq:total_weight}.}
\medskip
\small
\label{tab:notation}
\begin{tabular}{r p{6.5cm}}
\hline \T
$J_a, m_a$ & Angular momentum quantum numbers for the initial state. \\
$J_b, m_b$ & Same for the intermediate state. \\
$l_1, l_2$ & Orbital angular momentum in the primary and secondary breakup, respectively \\
$\lambda_b$ & The level populated in the intermediate system. Implicitly specifies $J_b$ and $l_2$.\\
$c$ & Decay channel specifying $\lbrace l_1, \lambda_b \rbrace$.\\
$\gamma_c$ & Reduced width amplitude for decay of the initial state through channel $c$. \\
$\gamma_{\lambda_b l_2}$ & Same for decay of the intermediate state. \\
$\Omega_1$ & Direction of the first emitted $\alpha$ in the rest frame of the initial state. \\
$\Omega_{23}$ & Direction of the second emitted $\alpha$ in the rest frame of the intermediate state. \\
$E_{23}$ & Relative energy between $\alpha_2$ and $\alpha_3$ \\
$\rho_1$ & $=k_1 a_1$, where $k_1$ is the wave number and $a_1$ is the channel radius for the primary breakup channel. \\
$\rho_{23}$ & Same for the secondary breakup channel.\\
$P_{l_1}, P_{l_2}$ & Penetrability for the primary and secondary breakup channels. \\
$\omega_{l_1}, \omega_{l_2}$ & Coulomb phase shifts. \\
$\phi_{l_1}, \phi_{l_2}$ & Hard-sphere phase shifts. \\
$E_{\lambda_b}$ & Level energy of $\lambda_b$ in the intermediate system. \\
$S_{l_2}, B_{l_2}$ & Shift function and boundary condition for the secondary breakup channel.   \B \\
\hline
\end{tabular}
\end{table}

\emph{Model I} has some desirable features: First, the amplitude is determined by standard $R$-matrix level parameters, which can be obtained from $\alpha\alpha$ scattering or from the analysis of $\beta$-delayed $\alpha$ spectra from the decay of $^8\mathrm{Li}$ and/or $^8\mathrm{B}$. Second, the model takes the identity of the $\alpha$ particles into account by treating them as bosons and symmetrising with respect to their labelling. Finally, it has been shown to fit the phase-space distributions of the $\alpha$ particles emitted by several excited states in ${^{12}\mathrm{C}}^*$, for instance the $J^\pi = 1^+$ state at $E_x = \SI{12.71}{\mega\electronvolt}$~\cite{Balamuth1974,Fynbo2003,Kirsebom2010}, the $2^+$ state at $E_x=\SI{16.11}{\mega\electronvolt}$~\cite{Schaefer1970,Laursen2016} and the $2^-$ state at $E_x = \SI{16.57}{\mega\electronvolt}$~\cite{Cockburn1970}, as well as observations of the $^3\mathrm{H}(^3\mathrm{H},nn\alpha)$ reaction at low energy~\cite{Brune2015}.

\subsection{Final state Coulomb interactions}
\label{sec:fsci}
\noindent \emph{Model I} takes final-state Coulomb interactions (FSCI) into account by including the penetrabilities for the primary and secondary breakup channels. This is only correct if the $\alpha_1 + {^8\mathrm{Be}}$ and $\alpha_2 + \alpha_3$ pairs are allowed to propagate to infinity in their relative coordinates. When the lifetime of the intermediate ${^8\mathrm{Be}}$ state becomes very short, however, that picture breaks down, and the treatment using only two-body Coulomb interactions becomes inaccurate, a point which has also been discussed by others~\cite{Cockburn1970,Fynbo2003}. A phenomenological approach to improving the description of FSCI has been proposed and tested against data from the decay of the $1^+$ state at $E_x = \SI{12.71}{\mega\electronvolt}$, which proceeds through the short-lived $2^+$ state at $E_x = \SI{3.0}{\mega\electronvolt}$ in $^8\mathrm{Be}$~\cite{Fynbo2003}. The idea is to let the fragments of the primary decay, initially separated by the channel radius $a_1$, propagate as usual out to some distance, $\tilde{r}$. At this point we replace the penetration factor of the $\alpha_1+{^8\mathrm{Be}}$ pair by the product of penetration factors for the $\alpha_1 + \alpha_2$ and $\alpha_1 + \alpha_3$ pairs. Formally, we make the following substitution in eq. \eqref{eq:single_amp},
\begin{align}
\label{eq:correction}
\frac{P_{l_1}}{\rho_1} \; \rightarrow \; \frac{P_{l_1}}{\rho_1} \Biggl[\frac{\tilde{\rho}_1}{\tilde{P}_{l_1}}\frac{\tilde{P}_{l_2}(E_{12})}{\tilde{\rho}_{12}}\frac{\tilde{P}_{l_2}(E_{13})}{\tilde{\rho}_{13}}\Biggr],
\end{align}
where the tilde functions are the usual $R$-matrix functions evaluated at $\tilde{r}$ and $E_{ij}$ is the relative energy between $\alpha_i$ and $\alpha_j$. In this way the Coulomb interactions of each $\alpha$ pair is treated symmetrically. This modified version of the sequential decay model is our \emph{Model II}. A slightly different modification of the penetration factor was made in \cite{Brune2015}. While their modification may give the correct behaviour for small $E_{12}$ and $E_{13}$, its interpretation in terms of transmission probabilities is not as clear as eq. \eqref{eq:correction}.

%It is not obvious that these considerations are relevant for the sequential Hoyle state decay; after all, the partial width of the $^8\mathrm{Be}$ ground state suggests it to be quite long-lived. It is true that the $^8\mathrm{Be}$ ground state is long-lived when it is populated on-resonance, but when the state is populated off-resonance, for instance if the ghost is populated, it has a much shorter lifetime, comparable to that of the first excited $2^+$ state (see Sec. \ref{sec:lifetime}). As an experiment we take a value $\tilde{r} = \SI{15}{\femto\meter}$ and apply the correction for FSCI to our sequential model. Calculation of the phase space distribution of the Hoyle state decay results in the plot shown in the right panel of \fref{fig:hoyle}. We note a relative suppression of the decay weight near the upper right corner of the Dalitz plot, corresponding to decays where the three $\alpha$ particles are emitted with equal energies. Intuitively this is a sensible result, since the FSCI would tend to suppress decays where any of the $\alpha$ particle pairs appear with a small relative energy. To quantify the effect, we calculate the relative intensity outside the peak, and the results are listed in \tref{tab:results}. The enhancement of decays with a large $E_{23}$ is around one order of magnitude, which shows that a proper description of the FSCI in three-body decays is required in order to make detailed predictions about the Hoyle state decay.

\subsection{Lifetime of the intermediate state}
\label{sec:lifetime}
\noindent By using a single $\tilde{r}$ throughout the entire phase space we implicitly assume that the lifetime of the intermediate system is constant and independent of the division of energy between the decay fragments. It has been shown that the lifetime of a nuclear resonance is in fact energy dependent and can be calculated from the resonant phase shift~\cite{Wigner1954,Smith1960,Smith1962,Baz1965}:
\begin{align}
\label{eq:lifetime}
\tau_2 = \hbar \frac{d\delta_2}{dE_{23}} + \frac{a_2}{v_{23}},
\end{align}
where $\delta_2$ is the $\alpha\alpha$ scattering phase shift, $a_2$ is the channel radius for the secondary breakup channel and $v_{23}$ is the relative velocity of the $\alpha$ particles emitted in the secondary breakup, which we approximate by its asymptotic value for $r \rightarrow \infty$. From this result we see that the lifetime is largest if the intermediate system is populated on-resonance, where the phase shift increases sharply. Off-resonance the lifetime is shorter, and it may even become negative.

%\noindent There is one obvious problem in using a single $\tilde{r}$ throughout the phase space, namely that the same correction is applied to decays regardless of whether the intermediate system is populated on-resonance or off-resonance. If the intermediate system is populated on-resonance, it will live long enough to travel $\simeq\SI{e6}{\femto\meter}$ before breaking up, whereas off-resonance it will only be able to travel a few \si{\femto\meter}. The use of an $\tilde{r} = \SI{15}{\femto\meter}$ thus becomes somewhat questionable. In the following we attempt to formulate a more systematic way of treating the FSCI.

\begin{figure}[h]
\centering
\includegraphics[width=0.9\columnwidth]{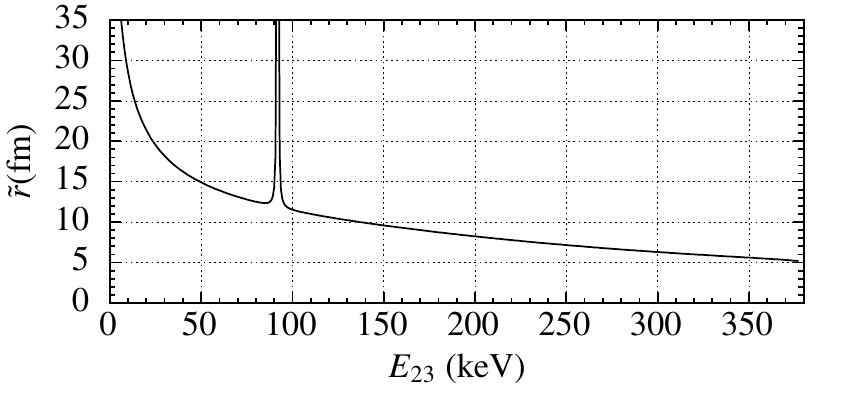}
\includegraphics[width=0.9\columnwidth]{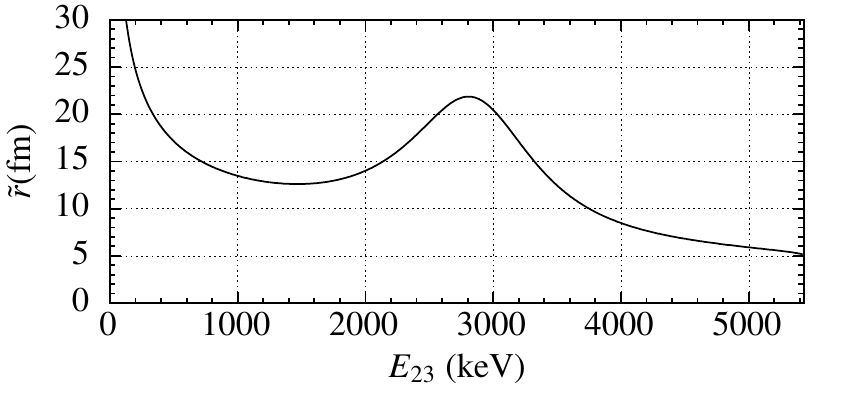}
\caption{Distance travelled by the fragments of the primary breakup before the breakup of the intermediate system plotted against the relative energy of the fragments in the secondary breakup; calculated from eq. \eqref{eq:rtilde} using the parameter values in \tref{tab:parameters}. The upper graph is relevant for the breakup of the Hoyle state through $^8\mathrm{Be(0^+)}$, while the lower graph shows the situation for breakup of the \SI{12.7}{\mega\electronvolt} $1^+$ state through $^8\mathrm{Be(2^+)}$.}
\label{fig:rtilde}
\end{figure}

We use the lifetime from eq. \eqref{eq:lifetime} and a simple, classical picture to estimate the distance between the primary fragments when the secondary breakup takes place: Suppose that the primary fragments are formed on the channel surface of the primary system, i.e. they are initially seperated by a distance $a_1$. The fragments now separate at a velocity, $v_1$, determined by their relative kinetic energy. Since at this point the fragments are tunnelling through the Coulomb barrier, the relative kinetic energy is, in a classical picture, not a well-defined quantity, but we assume it to be equal to the relative kinetic energy for $r \rightarrow \infty$. An estimate for $\tilde{r}$ in terms of the lifetime of the secondary resonance, $\tau_2$, then becomes
\begin{align}
\label{eq:rtilde}
\tilde{r} = a_1 + v_1 \tau_2 .
\end{align}
In order to get a feeling for the magnitude and behaviour of $\tilde{r}$ we look at two examples: The decay of the Hoyle state proceeding through the $0^+$ ground state in $^8\mathrm{Be}$ with a $Q$-value of \SI{379}{\kilo\electronvolt}, and the decay of the $1^+$ state at $E_x = \SI{12.71}{\mega\electronvolt}$, which proceeds through the $2^+$ first excited state in $^8\mathrm{Be}$ with a $Q$-value of \SI{5434}{\kilo\electronvolt}. We use the $R$-matrix parameters listed in \tref{tab:parameters}. The resulting value of $\tilde{r}$ is shown in \fref{fig:rtilde} as function of the internal energy in the intermediate system.
\begin{table}[htbp]
\centering
\caption{$R$-matrix parameters for the relevant levels in $^8\mathrm{Be}$. $E$ and $\Gamma_|obs|$ of the $0^+$ level are taken from \cite{Tilley2004} while $E$ and $\gamma^2$ of the $2^+$ level are taken from \cite{Bhattacharya2006}. The other figures were calculated using a channel radius of \SI{4.5}{\femto\meter} and standard $R$-matrix formulas~\cite{Lane1958}.}
\medskip
\begin{tabular}{c c c c}
\hline
$J^\pi$ & $E$ (\si{\kilo\electronvolt}) & $\Gamma_|obs|$ (\si{\kilo\electronvolt}) & $\gamma^2$ (\si{\kilo\electronvolt}) \T\B \\
\hline
$0^+$ & \num{91.84 \pm 0.04} & \num{5.57 \pm 0.25 e-3} & \num{830 \pm 38} \T \\
$2^+$ & \num{3129 \pm 6} & \num{1477 \pm 13} & \num{1075 \pm 9} \B\\
\hline
\end{tabular}
\label{tab:parameters}
\end{table}
For the Hoyle-state decay we see that for most values of $E_{23}$, the resulting value of $\tilde{r}$ is in fact quite small, and we should expect FSCI to have a pronounced effect on the breakup. If the intermediate system is populated on-resonance, however, it lives long enough to travel $\simeq\SI{e6}{\femto\meter}$ before breaking up. In this case we expect that the approximations of \emph{Model I} are very good. For the decay of the $1^+$ state $\tilde{r}$ show only small variations around an average of approximately \SI{15}{\femto\meter}. In previous studies \emph{Model II}, using a constant $\tilde{r}$ of around \SI{15}{\femto\meter}, has been shown to provide a reasonable fit to experimental data for the $1^+$ state~\cite{Fynbo2003,Refsgaard2016} and for the $2^+$ state at $E_x=\SI{16.11}{\mega\electronvolt}$, which also decays through the $2^+$ resonance in $^{8}\mathrm{Be}$~\cite{Laursen2016}\footnote{Due to a calculational error in Refs.~\cite{Fynbo2003} and \cite{Laursen2016} the value quoted in these references ($\tilde{r} = \SI{10}{\femto\meter}$) is too small. Better agreement with data is found for a somewhat larger value of $\tilde{r}$.}. It is remarkable that the simple estimate of eq. \eqref{eq:rtilde}, which does not include any adjustable parameters (except for the channel radii), is in agreement with the empirical values.

We conclude that for some decays \emph{Model II} is a good approximation, but also that we can not assume it to be generally applicable. Therefore we introduce \emph{Model III}, where the decay weight is calculated from eq.~\eqref{eq:total_weight} and the correction for FSCI in eq.~\eqref{eq:correction} is applied using a variable $\tilde{r}$, found from eq.~\eqref{eq:rtilde}. Based on the considerations in the preceding paragraph we expect \emph{Model III} to perform as well, or better, than \emph{Model II}.

\begin{table}[h]
\centering
\caption{Overview and description of the various models presented in the text.}
\label{tab:overview}
\medskip
\small
\begin{tabular}{r p{6.5cm}}
\hline 
\T \emph{Model I} &  Sequential $R$-matrix model. Symmetric with respect to exchange of any $\alpha$ pair. Decay weight calculated directly from eqs.~\eqref{eq:single_amp}--\eqref{eq:total_weight}. \\
\emph{Model II} & Similar to \emph{Model I}, but the change shown in eq.~\eqref{eq:correction} has been made in order to accomodate the finite lifetime and travel length of the intermediate fragment. \\
\emph{Model III} & Similar to \emph{Model II}, but with variable lifetime of the intermediate fragment. \B \\
\hline 
\end{tabular}
\end{table}

%One proceeding through $^8\mathrm{Be(0^+)}$, having $Q_{3\alpha}=\SI{379}{\kilo\electronvolt}$, and another decay, going through $^8\mathrm{Be(2^+)}$, with $Q_{3\alpha}=\SI{5434}{\kilo\electronvolt}$ (see \fref{fig:rtilde}). These correspond to the decay of the Hoyle state and the \SI{12.7}{\mega\electronvolt} $1^+$ state, respectively.

%\subsection{Comparison to data}
\subsection{The Dalitz plot}
\label{sec:comparison}
\noindent Often the Dalitz plot is used to represent the three-body final states that are observed in experiments and to visualise the predictions of theoretical models~\cite{Dalitz1953}. The coordinates of the plot are defined by
\begin{align}
\label{eq:Dalitz}
x = \frac{\sqrt{3}(E_1-E_3)}{Q} \quad \text{and} \quad y = \frac{2E_2 - E_1 - E_3}{Q},
\end{align}
where $E_i$ is the kinetic energy of the $i$th $\alpha$ particle in the rest frame of the decaying nucleus, ordered such that $E_1 > E_2 > E_3$, and $Q=\sum_{i}E_i$. All decays fulfilling energy and momentum conservation can be represented by a point inside the pie-wedge shaped region seen in Figs.~\ref{fig:hoyle} and \ref{fig:direct}. A point near the origin represents a decay where the available energy is shared equally between the three breakup fragments, while a point near the bottom right corner represents a decay with a small relative energy between the two lowest-energy fragments. Points near the top right corner of the plot represent decay where two of the fragments are emitted in opposite directions, leaving only very little energy to the third fragment.

\section{Sequential breakup}
\begin{figure}[htbp]
\centering
\includegraphics[width=0.48\columnwidth]{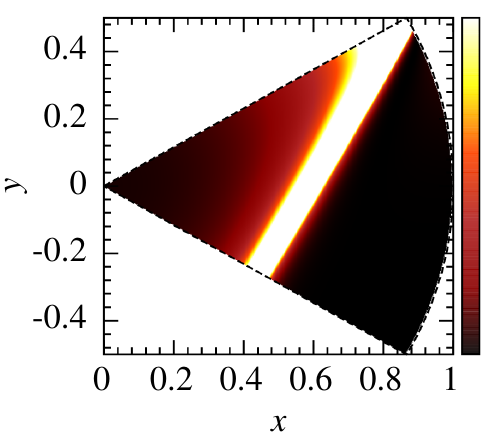}
\hspace*{\fill}
\includegraphics[width=0.48\columnwidth]{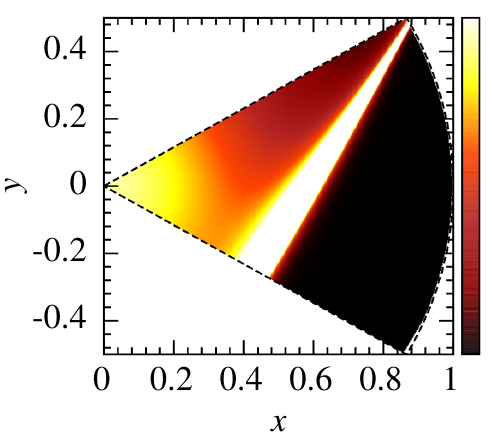}
\caption{Expected phase space distribution of $\alpha$ particles emitted in a sequential breakup of the Hoyle state as calculated with \emph{Model I} (left) and \emph{Model III} (right) plotted on a linear color scale. Note that the peak value is several orders of magnitude higher than the color scale limit.}
\label{fig:hoyle}
\end{figure}

\noindent Let us assume that the Hoyle state decays sequentially through the $0^+$ ground state of $^8\mathrm{Be}$. In \fref{fig:hoyle} we show the decay weight calculated with \emph{Model I} and \emph{Model III}, where we have chosen channel radii $a_1 = \SI{5.1}{\femto\meter}$ and $a_2 = \SI{4.5}{\femto\meter}$, corresponding to a value for the nucleon radius of $r_0 = \SI{1.42}{\femto\meter}$. We see that the weight is sharply peaked on a diagonal line corresponding to a relative energy between the lowest-energy alphas of $E_{23}=\SI{91.84}{\kilo\electronvolt}$. This line is usually interpreted as a signature of the sequential decay through $^8\mathrm{Be}(0^+)$, however, we also note that the models predict a low-intensity tail stretching from the diagonal line towards the apex of the Dalitz plot, \emph{Model III} most significantly so.

The intensity outside the ${^8\mathrm{Be}}$ ground-state peak is related to the so-called \emph{ghost anomaly}, which appears for nuclear levels near thresholds~\cite{Beckner1961,Barker1962}. In fact, if we consider a normalised $R$-matrix lineshape
\begin{align}
w(E) = \pi^{-1} \frac{\Gamma_\lambda/2}{\bigl[E_{\lambda} - E - \gamma^2 (S - B)\bigr]^2 + \bigl[\Gamma_\lambda/2\bigr]^2}
\end{align}
then we can approximate the area under a narrow peak as
\begin{align}
\int_{E_\lambda - \delta E}^{E_\lambda + \delta E} w(E)dE \,\simeq\, \biggl[1+\gamma^2\Bigl(\frac{dS}{dE}\Bigr)_{E_\lambda} \biggr]^{-1}.
\end{align}
From this expression we see that the peak area is dependent on both the reduced width and, through the derivative of the shift function, on the channel radius. For $a_2 = \SI{4.5}{\femto\meter}$ we find that only \SI{57 \pm 2}{\percent} of the ${^8\mathrm{Be}}$ ground-state strength appears in the observed ground state peak, the uncertainties coming from the quoted uncertainty on the partial width of the ground state. Since we are free to choose other values for the channel radius, it should be pointed out that the estimate is quite sensitive to this parameter. With the larger radius $a_2 = \SI{7.0}{\femto\meter}$ the peak area increases to \SI{86 \pm 1}{\percent}. The strength we see outside the peak in \fref{fig:hoyle} is the hint of a ghost anomaly, although heavily suppressed by Coulomb-barrier effects in the primary decay channel. To quantify how large a fraction of the decays we expect to observe outside the ground-state peak, we use a Monte-Carlo routine to integrate the decay weight over the region where $E_{23}>E_|gs| + \delta E$. The resulting fractional intensities are listed in \tref{tab:results} for a few values of $\delta E$. The values vary within \SI{\pm 10}{\percent} when the channel radii are varied between \SI{1.42}{\femto\meter} and \SI{2}{\femto\meter}. The same order of sensitivity is seen for variations of $\Gamma_|gs|$ within the experimental uncertainties.
\begin{table}[htbp]
\centering
\caption{Fractional intensity of decays with $E_{23} > E_|gs| + \delta E$, calculated using Monte-Carlo integration of the three models listed in \tref{tab:overview}. In \emph{Model II} we have used $\tilde{r}=\SI{16}{\femto\meter}$. Also shown are the values, $I_F$, obtained from a more sophisticated calculation, involving the solution of the Faddeev equations for the $3\alpha$ system~\cite{Ishikawa2017}.}
\label{tab:results}
\medskip
\small
\begin{tabular}{c c c c c}
\hline 
{$\delta E \;(\si{\kilo\electronvolt})$} & {$I_|I|$} & {$I_|II|$} & {$I_|III|$} & $I_F$ \T\B \\ 
\hline 
10 & \num{2.3e-4} & \num{8.8e-4} & \num{1.1e-2} & \num{5.2e-4} \T \\
20 & \num{1.1e-4} & \num{7.2e-4} & \num{8.4e-3} & \num{3.1e-4} \\
30 & \num{6.0e-5} & \num{5.6e-4} & \num{6.8e-3} & \num{2.2e-4} \\
50 & \num{1.7e-5} & \num{3.4e-4} & \num{4.0e-3} & \num{1.4e-4} \B \\
\hline
\end{tabular}
\end{table}

Looking at \fref{fig:hoyle} we note that the result of \emph{Model I} has a striking visual similarity with the prediction in Fig.~1(a) of Ref.~\cite{Ishikawa2014}, which was obtained by solving the Faddeev equations using $\alpha\alpha$ and $3\alpha$ interactions. In \tref{tab:results} we see that also quantitatively the three-body calculation is in closer accord with the results from \emph{Model I} and \emph{Model II} than those from \emph{Model III}. Experimentally, an upper limit for the fractional intensity for $\delta E \approx \SI{50}{\kilo\electronvolt}$ was recently found to be between \num{4.7e-4} and \num{4.2e-4}~\cite{Smith2017,DellAquila2017} at \SI{95}{\percent} C. L., which is consistent with \emph{Model I} and \emph{Model II}. It is remarkable that \emph{Model III} predicts a value which is an order of magnitude larger than the experimental upper limit.

\emph{Model III} clearly fails to describe the Hoyle state decay. This is surprising, since \emph{Model III}, which is based on the $R$-matrix framework and a physically motivated model of the three-body Coulomb interaction, reproduces values of $\tilde{r}$ found from the analysis of decay spectra of higher-lying states in $^{12}\mathrm{C}$, as discussed in Sec. \ref{sec:lifetime}. One major difference between the Hoyle state and the higher-lying states is that the Hoyle state sits behind a Coulomb barrier of around \SI{35}{\femto\meter}, while the barrier for the $1^+$ state at $E_x =\SI{12.71}{\mega\electronvolt}$ is only a few \si{\femto\meter} wide. As was mentioned at the introduction of \emph{Model III}, it treats all relative motion classically using asymptotic values of the kinetic energies. This approach is clearly problematic, in particular when the particles are moving inside classically forbidden regions, where the concept of velocity becomes ill-defined. Both theoretical and experimental investigations suggest that the effective velocity of a particle tunnelling through a wide barrier is, if anything, significantly larger than the asymptotic value~\cite{Winful2006}. Taking this into account we expect the values of $\tilde{r}$ shown in \fref{fig:rtilde} to be somewhat underestimated for the Hoyle-state decay. Using larger values of $\tilde{r}$ would tend to diminish the importance of three-body Coulomb interactions and to bring the results of our \emph{Model III} in better agreement with both theoretical end experimental results.

%It is well-known that the effective velocity of a particle tunnelling through a wide barrier can be significantly larger than the asymptotic value~\cite{Winful2006}

%This approach is clearly problematic, in particular when the particles are moving inside classically forbidden regions, where the concept of velocity becomes somewhat ill-defined. Both theoretical and experimental investigations suggest, however, that the effective velocity of a particle tunnelling through a wide barrier is, if anything, significantly larger than the asymptotic value~\cite{Winful2006}

\section{Direct breakup}
\label{sec:direct}
\begin{figure}[htbp]
\centering
\includegraphics[width=0.48\columnwidth]{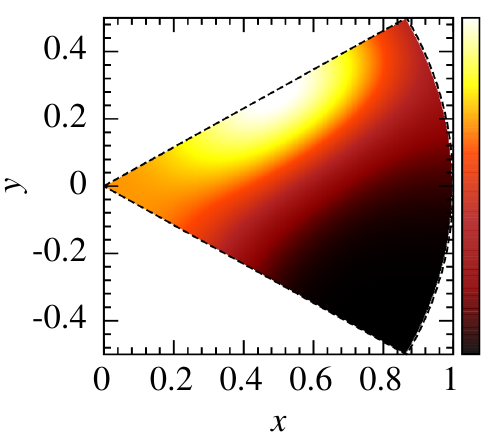}
\hspace*{\fill}
\includegraphics[width=0.48\columnwidth]{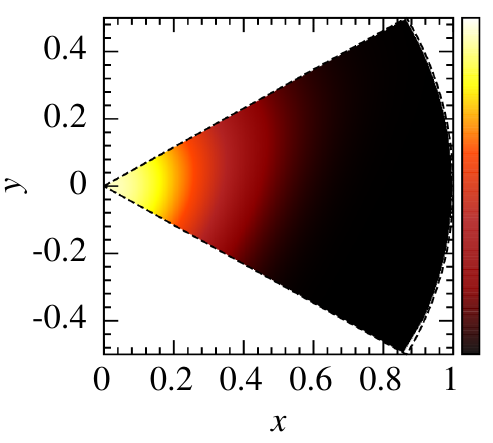}
\caption{Prediction of the phase space distribution for the direct breakup of the Hoyle state, calculated using \emph{Model I} (left) and \emph{Model III} (right).}
\label{fig:direct}
\end{figure}

\noindent Would it be possible to tweak the sequential model and make it predict the phase-space distribution of a direct decay? It is indeed possible to describe direct reactions in the $R$-matrix framework, but it requires the inclusion of infinitely many levels in the compound nucleus~\cite{Wigner1947,Lane1958}. In practice, however, what is most often done is to include a single \emph{background pole}; a very broad level at high excitation energy. Therefore, we attempt to calculate the phase space distribution of a direct breakup of the Hoyle state by replacing the $^8\mathrm{Be}$ ground state with a $0^+$ resonance at $E_|bg| = \SI{20}{\mega\electronvolt}$ and a width of $\Gamma_|bg| = \SI{200}{\mega\electronvolt}$. The result is shown in \fref{fig:direct}. Note that the distribution is not sensitive to our particular choice of $E_|bg|$ and $\Gamma_|bg|$, as long as the  level energy is far outside the range of energies that are relevant for the Hoyle state decay.
We see a relative suppression of the decay weight near the lower right corner, which represents decays with a small $E_{23}$. The main difference between \emph{Model I} and \emph{Model III} is that \emph{Model III} also predicts a suppression near the top right corner of the Dalitz plot. Intuitively this is a sensible result, since the FSCI would tend to suppress decays where any of the $\alpha$-particle pairs appear with a small relative energy. It seems that we should expect a hypothetical direct decay of the Hoyle state to show up as a sharp peak near the apex of the Dalitz plot. We believe that our model for the direct decay is more accurate than the na\"{i}ve estimates using a uniform phase space decay used in~\cite{Raduta2011,Manfredi2012,Kirsebom2012,Rana2013,Itoh2014,Smith2017,DellAquila2017}, since we, at least in some approximation, include Coulomb interactions between each $\alpha$ particle in the final state.

An alternative way to predict the phase-space distribution of a direct $3\alpha$ decay is presented in Ref.~\cite{Smith2017b}, where a uniform phase-space decay is combined with a Coulomb-barrier transmission probability calculated using the WKB approximation in hyperspherical coordinates~\cite{Garrido2005}. The obtained phase space distribution is very similar to the result of our \emph{Model III}. The transmission probability derived in \cite{Garrido2005} can be calculated for both direct and sequential breakup, but the method does not predict which approximation is the most suitable. It is one strength of \emph{Model III} that the penetration factor, through the variable $\tilde{r}$, can be modified continuously between the sequential and direct limits, and that each part of phase space can be treated in the appropriate approximation.
%The transmission probability derived in \cite{Garrido2005} can be calculated for both direct and sequential breakup, but the method does not predict which approximation is the most appropriate. The results of Sec. \ref{sec:lifetime}, which \emph{Model III} has built-in, provide guidance as to which part of the phase-space should be treated in the sequential and direct picture, respectively. 

%We see a relative suppression of the decay weight near the upper right corner of the Dalitz plot, corresponding to decays where the three $\alpha$ particles are emitted with equal energies. Intuitively this is a sensible result, since the FSCI would tend to suppress decays where any of the $\alpha$ particle pairs appear with a small relative energy.

\subsection{Interference between decay channels}
\noindent We know from experiment that the Hoyle state has a sizeable sequential branch ($\simeq\SI{100}{\percent}$). Therefore we will never observe a pure, direct decay, but only a mixture of sequential and direct decay, which means that we need to revise the single-level approximation of eq. \eqref{eq:single_level}. A procedure for treating multiple levels in the intermediate system of sequential reactions has been proposed in \cite{Barker1967,Barker1988}, and we replace eq. \eqref{eq:single_level} with
\begin{align}
\label{eq:multi_level}
F_c(E_{23}) =\sum_{\mu_b}\bigl[ A_{\lambda_b \mu_b} \gamma_{\mu_b l_2} \bigr] \bigl(2P_{l_2} / \rho_{23}\bigr)^{\frac{1}{2}}\exp\bigl[i(\omega_{l_2} - \phi_{l_2})\bigr],
\end{align}
where $A_{\lambda_b \mu_b}$ is the level matrix for the intermediate system, defined by the relation
\begin{align}
(A^{-1})_{\lambda_b \mu_b} = (E_{\lambda_b}-E_{23})\delta_{\lambda_b \mu_b} - \sum_{c}(S_c - B_c + iP_c) \gamma_{\lambda_b c}\gamma_{\mu_b c} .
\end{align}
With this modification it is straightforward to calculate the theoretical phase-space distribution for various mixtures of sequential and direct decay.

\begin{figure}[htbp]
\centering
\includegraphics[width=0.92\columnwidth]{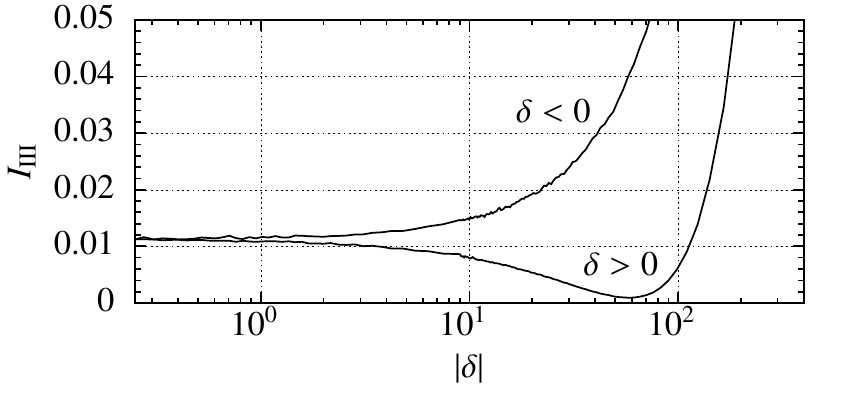}
\caption{Fraction of Hoyle state decays with $E_{23} > E_{gs} + \SI{10}{\kilo\electronvolt}$ calculated using \emph{Model III}. Two levels have been included in the intermediate $^8\mathrm{Be}$ system: The narrow ground state and a broad background pole. The mixing ratio, $\delta$, is defined in the text.}
\label{fig:interference}
\end{figure}

The reduced width amplitude, $\gamma_c$, of eq. \eqref{eq:single_amp} is the parameter which specifies the contribution of each decay channel (see also \tref{tab:notation}). If we consider the possibility that the Hoyle state can decay through both the ground state of $^8\mathrm{Be}$ and through the background pole introduced in Sec. \ref{sec:direct} we need two reduced width amplitudes, which we label $\gamma_|gs|$ and $\gamma_|bg|$. The mixing ratio $\delta = \gamma_|bg| / \gamma_|gs|$ determines the phase-space distribution of the decay products. In order to make a quantitative assessment of the effect of interference between the two decay channels we evaluate the decay weight using \emph{Model III} and find the fractional intensity for decays with $E_{23} > E_|gs| + \SI{10}{\kilo\electronvolt}$. The result is shown in \fref{fig:interference}.
Intuitively we should expect interference effects to be important only in the region of the Dalitz plot where both decay channels have an appreciable amplitude, which, judging from Figs. \ref{fig:hoyle} and \ref{fig:direct}, is near the apex of the Dalitz plot. The sign of $\delta$ determines whether the amplitudes in this part of the plot interfere constructively or destructively. It is clear from \fref{fig:interference} that destructive interference occur for $\delta\simeq +60$, where the fraction of Hoyle state decays with $E_{23} > E_|gs| + \SI{10}{\kilo\electronvolt}$ is diminished by an order of magnitude.

\section{Conclusion}
\noindent We have presented a schematic model to describe the effect of Coulomb interactions in the $3\alpha$ continuum and combined it with a well-established $R$-matrix formalism for sequential processes. We applied the model to the $3\alpha$ breakup of the Hoyle state in $^{12}\mathrm{C}$ and attempted to predict the phase-space distribution of the emitted $\alpha$ particles in a purely sequential decay. We observed considerable strength outside the $^8\mathrm{Be}$ ground-state peak, and our results suggest that the current experimental limit on the direct decay branch is very close to the point where we should start to observe this strength. The spectrum was seen to be quite sensitive to the way in which final-state Coulomb interactions are taken into account, and we expect that a careful measurement of the Hoyle-state decay will provide information on how to effectively treat three-body Coulomb interactions. We also presented a model which we believe contains the most important physics for direct three-body decay, as opposed to the simplistic assumptions of uniform phase-space decays, colinear decays etc., which appear in the literature. The Dalitz plot of the direct decay model showed an intensity peak near the origin, corresponding to decays with equal sharing of energy between the three $\alpha$ particles. Finally we showed that interference between the sequential and a possible direct decay channel could significantly alter the decay spectrum of the Hoyle state.

\section*{Acknowledgements}
\noindent We thank Prof. Souichi Ishikawa for making his results available to us. OSK acknowledges support from the Villum Foundation. We also acknowledge financial support from the European Research Council under ERC starting grant LOBENA, No.\ 307447.

%\section*{References}
\bibliography{bibliography}
\end{document}